\documentclass{IET-Conf-Paper}

\usepackage{amsmath,amssymb,amsfonts}
\usepackage{algorithmic}
\usepackage{graphicx}
\usepackage{textcomp}
\usepackage{xcolor}
\usepackage{stfloats}
\usepackage[caption=false]{subfig}

\usepackage{multicol}
\usepackage{booktabs}
\usepackage{url}
\usepackage[hidelinks]{hyperref}

\usepackage{cleveref}
\crefname{equation}{}{}
\Crefname{equation}{Equation}{Equations}
\crefname{figure}{Fig.}{Figs.}
\crefname{table}{Table}{Tables}
\crefname{section}{Section}{Sections}

\begin{document}

\title{TRAVELLING-WAVE-BASED PROTECTION OF DC HUBS IN BIPOLAR HVDC GRIDS}

\author{Rahul Rane\ad{1}\corr, Victor Daniel Reyes Dreke\ad{1}, Wilhelm Winter\ad{1,2}, Aleksandra Leki{\'c}\ad{1}}

\address{\add{1}{Intelligent Electrical Power Grids, Delft University of Technology, Delft, The Netherlands}
\add{2}{TenneT TSO GmbH, Bayreuth, Germany}
\email{r.m.rane@tudelft.nl}}

\keywords{Dedicated metallic return (DMR), fault detection, fault classification, wavelet transform}

\begin{abstract}
This paper presents a stationary wavelet transform (SWT)-based method for DC fault detection and classification in a three-terminal bipolar modular multilevel converter (MMC)-based HVDC system with a dedicated metallic return (DMR). The system incorporates DC switching stations (DCSSs) with interconnected DC circuit breakers (DCCBs). The method uses line-mode voltage wavelet detail energy for fault detection and wavelet detail energy derived from pole voltages and DMR current for fault classification, distinguishing pole-to-pole, pole-to-ground, and pole-to-DMR faults. Its performance is evaluated using a real-time hardware-in-the-loop (HIL) platform integrating an real-time digital simulator (RTDS) with embedded processing hardware. Busbar and cable faults with varying fault types, resistances, and locations are investigated. The results demonstrate the feasibility of real-time implementation and characterize the effects of fault location and travelling-wave propagation on transient energy and protection response time. The effectiveness of pole-voltage and DMR-current detail energy features in distinguishing fault types is also evaluated.
\end{abstract}

\maketitle

\section{Introduction}
The increasing integration of offshore wind power is driving the development of high-voltage direct-current (HVDC) transmission networks. In Europe, an additional 487~GW of offshore wind capacity has been identified in long-term development plans, highlighting the need for expanded HVDC infrastructure \cite{entsoe2024ondp,entsoe2024ondpNorthernSeas}. While existing HVDC connections are predominantly point-to-point, future systems are expected to evolve towards meshed multi-terminal HVDC (MTDC) networks to improve connectivity and transmission flexibility. This transition introduces new protection challenges, as AC-side protection alone cannot selectively isolate DC faults. Consequently, DC circuit breakers (DCCBs) and suitable DC fault-protection algorithms are essential for reliable MTDC operation.

Within meshed MTDC networks, DC hubs serve as interconnection points for multiple transmission corridors and converter terminals, enabling flexible power transfer between network branches. Each hub requires switching and protection capabilities to isolate faulted sections while maintaining connections to healthy parts of the network. DCCBs are designed to interrupt high DC fault currents and isolate faults, thereby protecting the HVDC system and supporting reliable operation \cite{cigre2022hvdcBreakers,belda2025hvdcSwitchgear}. The design and application of HVDC switching equipment must account for equipment performance and system-level protection requirements \cite{cigre2017hvdcSwitching}. However, achieving selective fault clearing across increasingly complex MTDC networks requires effective coordination between protection layers. The DC switching station (DCSS) provides the necessary switching and protection capabilities through an arrangement of DCCBs. This arrangement is particularly important for busbar faults, which can affect multiple connected circuits. Without suitable breaker coordination, clearing such faults may require disconnecting healthy connections, reducing network availability. Interconnected DCCBs support selective fault isolation and help limit the impact of faults on the wider MTDC network \cite{khan2024performance}. In parallel, environmental concerns associated with conventional switchgear have encouraged the development and adoption of alternatives to SF$_6$-based equipment \cite{entsoe2020sf6}, reflecting the broader evolution of high-voltage switching technologies.

HVDC protection methods can broadly be classified into single-ended and multi-ended schemes, depending on whether they rely on local measurements or exchange information between terminals \cite{jahn2017survey}. Single-ended methods include derivative-based techniques, travelling-wave analysis, and voltage-threshold protection \cite{naidoo2004hvdc,liu2023robust,kunlun2011study}. These methods can provide fast detection without inter-terminal communication, but their performance may be affected by measurement noise, fault resistance, line characteristics, and threshold selection. Multi-ended methods, including current differential protection and directional comparison, use measurements from multiple terminals to improve selectivity \cite{kobet2010justifying,he2013research}. However, their reliance on communication can introduce delays and additional implementation requirements, particularly in geographically extensive networks \cite{farkhani2024comprehensive}. These trade-offs motivate protection methods that provide timely fault detection and reliable classification under varying fault conditions and network configurations.

\begin{figure*}[b]
    \centering
    \subfloat[\label{fig:topology}]{\includegraphics[width=0.68\linewidth]{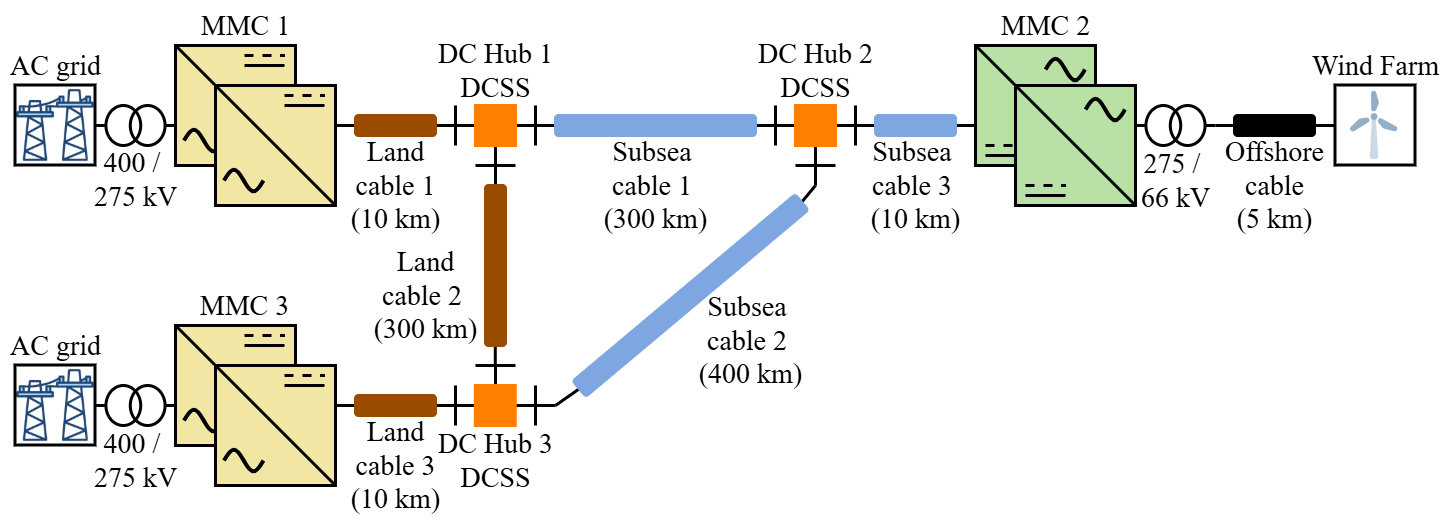}} \qquad
    \subfloat[\label{fig:dcss}]{\includegraphics[width=0.28\linewidth]{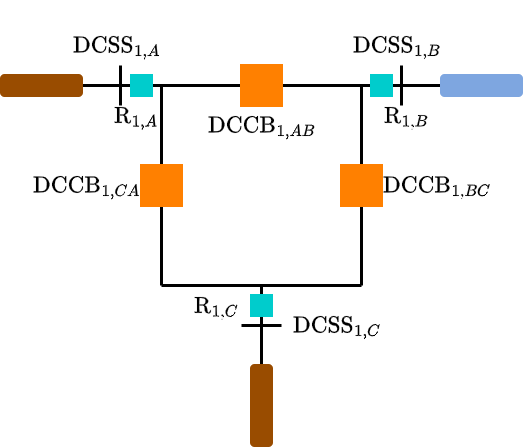}}
    \caption{Three-terminal multi-terminal HVDC test system: (a) overall system topology; (b) DCSS topology for DC hub 1 and interconnected DCCB arrangement for each pole.}
    \label{fig:test_system}
\end{figure*}

Signal-processing techniques can extract transient features from measured voltages and currents to support DC fault detection and classification \cite{german2018signal,uddin2018recent}. In particular, the wavelet transform (WT) provides time-frequency analysis suited to the non-stationary transients associated with DC faults \cite{de2011wavelet,psaras2020real}. Other approaches include Fourier-based transforms \cite{satpathi2017short}, the Stockwell transform \cite{li2020improved}, and the Hilbert--Huang transform \cite{albernaz2020distance}. Wavelet-based methods have also been investigated for travelling-wave and transient-based protection \cite{nanayakkara2012traveling,xiang2018transient,liu2023robust}. The stationary wavelet transform (SWT) avoids downsampling, preserving the temporal alignment of coefficients across decomposition levels \cite{liu2007stationary}. This property makes SWT suitable for extracting time-localized fault features from voltage and current measurements. Nevertheless, practical protection performance also depends on feature selection, detection and classification logic, network configuration, and operating conditions. Evaluation across different fault scenarios and real-time implementation constraints is therefore important.

Building on this background, this paper implements and evaluates an SWT-based fault-detection and classification method in a three-terminal bipolar modular multilevel converter (MMC)-based MTDC system with a dedicated metallic return (DMR). The system comprises DC hubs with DCSSs and interconnected DCCB arrangements. The method uses line-mode voltage detail energy for fault detection and wavelet detail energy derived from positive-pole voltage, negative-pole voltage, and DMR current for fault classification. The algorithm is implemented on embedded processing hardware and evaluated using a real-time hardware-in-the-loop (HIL) platform integrated with a real-time digital simulator (RTDS). Busbar and cable faults with varying fault types, resistances, and locations are investigated to assess detection and classification performance, transient-energy characteristics, and protection response time. The contribution of this study is the real-time implementation and systematic evaluation of the method within the considered DC hub protection configuration, rather than the introduction of a new wavelet transform.

The remainder of this paper is organized as follows. \cref{sec:test_system} describes the three-terminal bipolar MMC-HVDC test system and its DCSS arrangement. \cref{sec:fault_detection} presents the proposed SWT-based fault-detection and classification methods. \cref{sec:hil_results} describes the real-time HIL setup and discusses the results. Finally, \cref{sec:conclusion} concludes the paper.

\section{MTDC Test System Topology and DCCB Arrangement} \label{sec:test_system}
The three-terminal MMC-based multi-terminal HVDC system is implemented on RTDS for HIL protection studies. As shown in \cref{fig:test_system}, the system comprises two onshore converters connected to 400~kV AC grids and one offshore converter interfacing with an aggregated 1~GW-per-pole offshore wind farm through a 66~kV collection network and a 275/66~kV transformer. The onshore converters are connected to their respective AC grids through 400/275~kV transformers. The rated parameters of the onshore and offshore MMCs are adopted from \cite{rane2026real}. The bipolar DC network operates at $\pm 525$~kV and consists of three DC hubs interconnected by land and subsea cables. The link between hubs 1 and 2 comprises 300~km of subsea cable, whereas the links between hubs 2 and 3 and hubs 1 and 3 comprise 400~km of subsea cable and 300~km of land cable, respectively. MMCs 1 and 3 connect to hubs 1 and 3 through 10~km land cables, respectively, while MMC 2 connects to hub 2 through a 10~km subsea cable. All cables are represented using frequency-dependent phase-domain models that account for the positive pole, negative pole, and DMR, with cable parameters adopted from \cite{karmokar2025enhanced}.

For DC fault isolation, each hub incorporates a DCSS with an interconnected DCCB arrangement for each pole, as illustrated in \cref{fig:dcss}. Three relays, $R_A$, $R_B$, and $R_C$, monitor the positive-pole, negative-pole, and DMR voltages with respect to ground. The test system represents a baseline DCSS configuration in which each protected section is connected to a single cable. Although practical DC hubs may accommodate multiple cable connections per section \cite{khan2024performance}, the single-connection arrangement is adopted here to establish a baseline configuration and limit the complexity of the study.

\begin{figure}[t]
    \centering
    \includegraphics[width=0.75\linewidth]{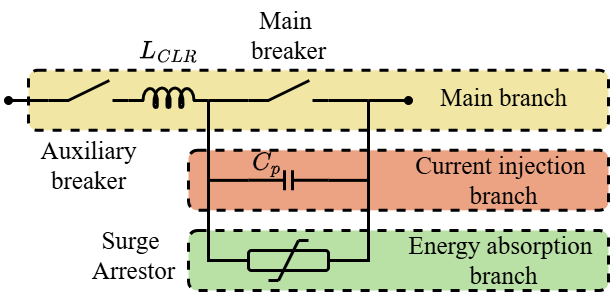}
    \caption{Average DC circuit breaker model based on \cite{rane2026real,khan2024performance,popov2023mechanicalDCCB}}
    \label{fig:dccb_model}
\end{figure}

This interconnected breaker arrangement is intended to isolate the faulted section while preserving healthy connections. In contrast, configurations with DCCBs only at transmission-line terminals may require the disconnection of multiple circuits to clear a busbar fault, potentially interrupting healthy connections \cite{khan2024performance}. To evaluate the protection arrangement, DCCBs are represented using average-value models that capture the principal electrical behaviour during current interruption without resolving detailed switching dynamics. The breaker model and its parameters are shown in \cref{fig:dccb_model} and listed in \cref{tab:dccb_parameters}, respectively.

\begin{table}[t]
    \centering
    \caption{Parameters of the average DCCB model}
    \label{tab:dccb_parameters}
    \begin{tabular}{lcc}
    \hline
    \textbf{Parameter} & \textbf{Symbol} & \textbf{Value} \\
    \hline
    Current limiting reactor & $L_{\mathrm{CLR}}$ & 250 mH \\
    Main breaker capability & $I_{\mathrm{main}}$ & 16 kA \\
    Auxiliary breaker capability & $I_{\mathrm{aux}}$ & 0.001 kA \\
    Parallel capacitor & $C_{\mathrm{p}}$ & 1.6 $\mu$F \\
    Arrester discharge current & $I_{\mathrm{SA}}$ & 20 kA \\
    Arrester discharge voltage & $V_{\mathrm{SA}}$ & 865 kV \\
    Arrester decay time constant & $\tau_{\mathrm{SA}}$ & 10 s \\
    \hline
    \end{tabular}
\end{table}

The MMC control strategies are selected according to the roles of the respective terminals. MMC~1 regulates the DC voltage and reactive power, MMC~2 operates in active- and reactive-power ($P$--$Q$) control mode, and MMC~3 employs grid-forming voltage--frequency ($V$--$f$) control to establish the offshore AC network. Capacitor-energy balancing and circulating-current suppression are also incorporated into the converter controls.

Finally, the complete RTDS model is evaluated under normal operating conditions and a range of DC fault scenarios. During these tests, the voltage measurements acquired by the DCSS relays are processed by the SWT-based protection method described in \cref{sec:fault_detection}. Detection accuracy, fault-classification performance, and response latency are assessed to evaluate the method's real-time implementation in the specified DC hub configuration.

\section{Fault Detection and Classification} \label{sec:fault_detection}
The proposed protection method comprises two stages: fault detection using SWT detail-coefficient energy and fault classification based on the detected involvement of the positive pole, negative pole, and DMR. The detection stage identifies fault-induced voltage transients, while the classification stage uses conductor-specific detection results to determine the fault type. Both stages are implemented at the respective relays to support fault detection, classification, and identification of the affected DCSS section. The following subsections describe the SWT-based transient analysis, the associated detection and classification algorithms, and the definition of detection time used to evaluate the protection response.

\subsection{SWT for Travelling-Wave Analysis} \label{subsec:wavelet_transforms}
The SWT decomposes a discrete-time signal into approximation and detail coefficients using low-pass and high-pass analysis filters, respectively. Unlike decimated wavelet transforms, the SWT preserves the original sampling rate by inserting zeros between filter coefficients at successive decomposition levels rather than downsampling the filtered signals. This property maintains the temporal alignment between the input signal and its wavelet coefficients, making the SWT suitable for analysing rapid transients associated with fault-induced travelling waves in DC networks.

With $A_0[n]=x[n]$, the approximation and detail coefficients at decomposition level $j$ are given by
\begin{subequations}
\begin{align}
    A_j[n] &= \sum_{k=0}^{N-1}h[k]A_{j-1}[n-2^{j-1}k],\\
    D_j[n] &= \sum_{k=0}^{N-1}g[k]A_{j-1}[n-2^{j-1}k],
\end{align}
\end{subequations}
where $h[k]$ and $g[k]$ are the low-pass and high-pass analysis filters, respectively, $N$ is the filter length, and $n$ is the sampling index. The resulting detail coefficients capture signal variations at the corresponding decomposition level. Their temporal alignment and preserved sample count facilitate moving-average energy analysis of fault-induced transients. Accordingly, the SWT provides the basis for both the line-mode voltage detection algorithm and the conductor-specific detection processes used for fault classification.

\subsection{Fault Detection Using Moving-Average Wavelet Energy} \label{subsec:fault_detection_energy}
To extract the voltage transient used for detection, the measured positive-pole and negative-pole voltages relative to ground are denoted by $v_{\mathrm{P}}[n]$ and $v_{\mathrm{N}}[n]$, respectively. These voltages are transformed into line-mode and zero-mode components as
\begin{equation}
    \begin{bmatrix}
        v_l[n]\\
        v_0[n]
    \end{bmatrix}
    =
    \frac{1}{2}
    \begin{bmatrix}
        1 & -1\\
        1 & 1
    \end{bmatrix}
    \begin{bmatrix}
        v_{\mathrm{P}}[n]\\
        v_{\mathrm{N}}[n]
    \end{bmatrix},
    \label{eq:mode_voltage_transformation}
\end{equation}
where $v_l[n]$ and $v_0[n]$ denote the line-mode and zero-mode voltages, respectively. The line-mode voltage $v_l[n]$ is selected as the input to the SWT-based fault-detection algorithm.

Using the SWT coefficients introduced in the previous subsection, let $D_{L,v_l}[n]$ denote the level-$L$ detail coefficient obtained from the line-mode voltage. The detection algorithm calculates the moving-average detail energy over a window of $M$ samples as:
\begin{equation}
    E_{L,v_l}[n]=\frac{1}{M}\sum_{i=0}^{M-1}
    \left|D_{L,v_l}[n-i]\right|^2.
    \label{eq:wavelet_energy}
\end{equation}
This energy quantifies the recent transient activity in the line-mode voltage. The instantaneous fault-detection decision $f_{v_l}[n]$ is then obtained by comparing the energy with a predefined threshold $E_{\mathrm{threshold},v_l}$:
\begin{equation}
    f_{v_l}[n]=
    \begin{cases}
        1, & E_{L,v_l}[n]>E_{\mathrm{threshold},v_l},\\
        0, & \text{otherwise}.
    \end{cases}
    \label{eq:instantaneous_fault_detection}
\end{equation}
Once the detection threshold is exceeded, the trip signal remains asserted even if the energy subsequently falls below the threshold. It is cleared only by an explicit reset. Thus, the detection stage provides a persistent fault indication based on the transient energy of the line-mode voltage. To determine which conductors are involved in the detected fault, the protection method additionally evaluates the positive-pole and negative-pole voltages and the DMR current, as described in the following subsection.

\subsection{Fault Classification} \label{subsec:fault_classification}
Following fault detection, the fault type is classified based on the involvement of the positive and negative poles and the DMR. At each relay, separate SWT-based detection algorithms process the positive-pole voltage $v_{\mathrm{P}}[n]$, negative-pole voltage $v_{\mathrm{N}}[n]$, and DMR current $i_{\mathrm{DMR}}[n]$. Their instantaneous binary detection outputs are denoted by $f_{v_{\mathrm{P}}}[n]$, $f_{v_{\mathrm{N}}}[n]$, and $f_{i_{\mathrm{DMR}}}[n]$, respectively. Each output indicates whether the moving-average SWT detail energy of the corresponding signal exceeds its predefined threshold. Similar to \cref{eq:instantaneous_fault_detection}, a value of $1$ indicates detected transient activity, whereas $0$ indicates that the threshold has not been exceeded. Together, these conductor-specific indicators provide the evidence used to distinguish among the defined fault types.

To retain this evidence as it becomes available, the three detection outputs are latched independently until reset. The fault class is determined from their latched states, $f_{v_{\mathrm{P}}}[n]$, $f_{v_{\mathrm{N}}}[n]$, and $f_{i_{\mathrm{DMR}}}[n]$, according to:
\begin{equation}
    C[n]=
    \begin{cases}
        \mathrm{Unknown}, & (1,1,1),\\
        \mathrm{P\text{-}N}, & (1,1,0),\\
        \mathrm{P\text{-}DMR}, & (1,0,1),\\
        \mathrm{N\text{-}DMR}, & (0,1,1),\\
        \mathrm{P\text{-}G}, & (1,0,0),\\
        \mathrm{N\text{-}G}, & (0,1,0),\\
        \mathrm{None}, & \text{otherwise},
    \end{cases}
    \label{eq:fault_classification}
\end{equation}
where the entries in each tuple correspond, in order, to $f_{v_{\mathrm{P}}}[n]$, $f_{v_{\mathrm{N}}}[n]$, and $f_{i_{\mathrm{DMR}}}[n]$. Here, $\mathrm{P\text{-}N}$, $\mathrm{P\text{-}DMR}$, and $\mathrm{N\text{-}DMR}$ denote positive-pole-to-negative-pole, positive-pole-to-DMR, and negative-pole-to-DMR faults, respectively. The classes $\mathrm{P\text{-}G}$ and $\mathrm{N\text{-}G}$ denote positive-pole-to-ground and negative-pole-to-ground faults. The $\mathrm{Unknown}$ class is assigned when all three detection indicators are asserted, as this combination does not uniquely identify one of the defined fault types. The $\mathrm{None}$ class indicates that no conductor-specific detection indicator is asserted.

Because the indicators are latched independently, the classification can be updated as additional conductor-specific evidence becomes available. For example, if $f_{v_{\mathrm{P}}}[n]$ is initially asserted while the other indicators remain inactive, the fault is classified as $\mathrm{P\text{-}G}$. If $f_{i_{\mathrm{DMR}}}[n]$ is subsequently asserted while $f_{v_{\mathrm{P}}}[n]$ remains active and $f_{v_{\mathrm{N}}}[n]$ remains inactive, the classification changes to $\mathrm{P\text{-}DMR}$. The resulting class is maintained according to the latched indicators until reset. This procedure is applied independently at each relay to determine the fault type and support identification of the affected DCSS section.

This proposed method is evaluated in terms of its wavelet energy response, fault detection time, and classification performance across different fault types, locations, and resistances. These aspects are investigated through real-time HIL testing, as described in the following section.

\section{Real-Time HIL Validation and Results} \label{sec:hil_results}
This section evaluates the proposed SWT-based protection method using a real-time HIL platform. The test configuration and SWT parameters are described first, followed by the assessment of cable and busbar faults under different fault types, resistances, and locations. The results address time-domain responses, wavelet-energy characteristics, detection time, and fault classification performance.

\subsection{HIL Test Configuration and SWT Parameters} \label{subsec:hil_setup_parameters}
\begin{figure}[htbp]
    \centering
    \includegraphics[width=\linewidth]{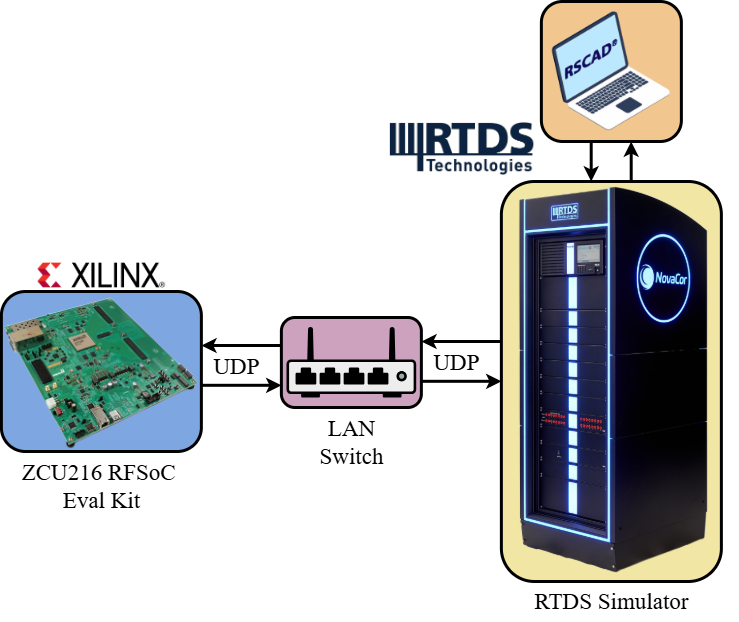}
    \caption{Protection HIL platform integrating the RTDS-based HVDC network simulation with the external hardware executing the proposed SWT-based protection algorithms.}
    \label{fig:hil}
\end{figure}

The proposed protection method is evaluated using the real-time HIL platform shown in \cref{fig:hil}. The RTDS emulates the three-terminal MMC-based MTDC system, while the protection algorithms execute on an AMD Zynq UltraScale+ RFSoC ZCU216 evaluation kit \cite{prosecco2025d41}. The SWT-based detection and classification algorithms run on the board's ARM Cortex-A53 processor. The RTDS and protection hardware exchange measurement data and protection signals over a 1-GbE UDP connection at a sampling rate of $50~\mathrm{kHz}$. The RTDS transmits the measured DC voltages and DMR current to the protection hardware, which processes the signals and returns detection and trip outputs to the RTDS to operate the simulated DCCBs.

The test matrix comprises five fault types (P-N, P-G, N-G, P-DMR, and N-DMR), five fault locations (the busbar at the DC hub 1 end of subsea cable 1 and four locations at $20\%$, $40\%$, $60\%$, and $80\%$ of the cable length measured from DC hub 1), and three fault resistances ($0.001~\Omega$, $10~\Omega$, and $100~\Omega$). The resulting test matrix comprises 75 fault cases.

The SWT employs the Daubechies-2 wavelet (db2) with a decomposition level of $L=4$. The moving-average energy window contains $M=20$ samples, corresponding to a duration of $400~\mu\mathrm{s}$ at the selected sampling rate. The detection thresholds are predefined for the respective measured signals. Once asserted, the detection outputs remain latched until an explicit reset is received.

\subsection{Cable Fault Response} \label{subsec:cable_fault_results}
Cable-fault performance is assessed under variations in fault type, resistance, and location. The time-domain responses are examined to characterise the voltage and current transients associated with faults involving and not involving the DMR. The influence of fault location on wavelet energy and detection time is subsequently evaluated across the cable-fault test cases.

\subsubsection{Fault Transients and Protection Response} \label{subsubsec:cable_waveforms}
Two representative faults on subsea cable 1 are analysed: a P-DMR fault at $80\%$ of the cable length measured from DC hub 1 and a P-G fault at $20\%$. Both faults are applied at $t=1~\mathrm{ms}$ with a fault resistance of $10~\Omega$. The P-DMR case illustrates the current response to a DMR fault, while the P-G case provides a comparison with a fault that does not involve the DMR.

\begin{figure}[htbp]
    \centering
    \includegraphics[width=\linewidth]{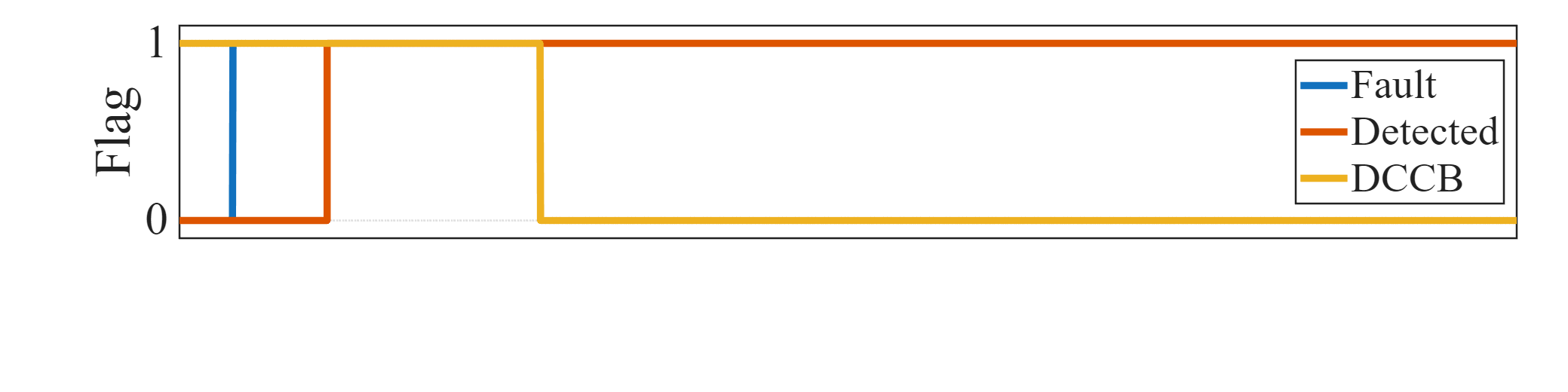} \\ [-0.5cm]
    \includegraphics[width=\linewidth]{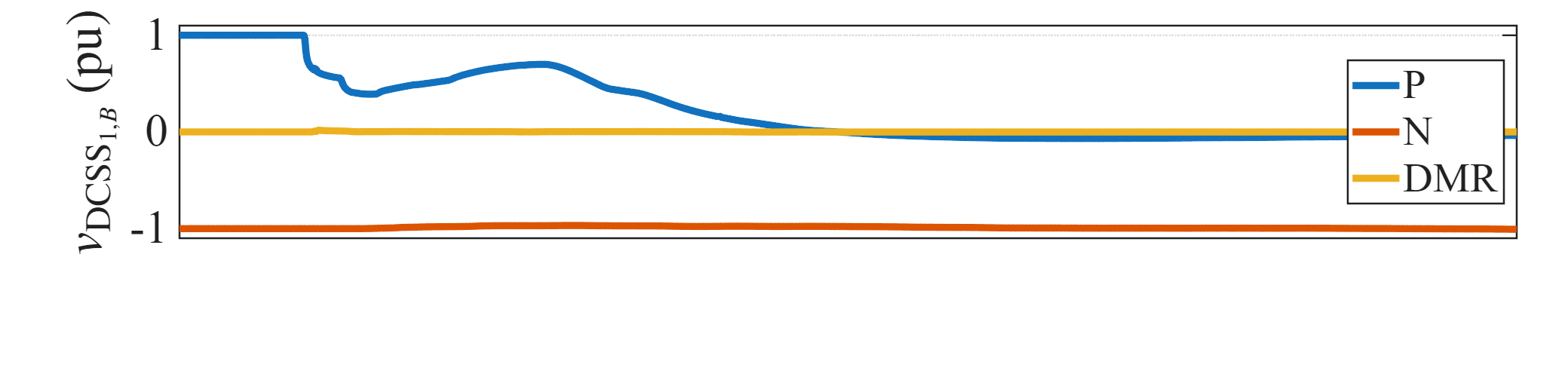} \\ [-0.5cm]
    \includegraphics[width=\linewidth]{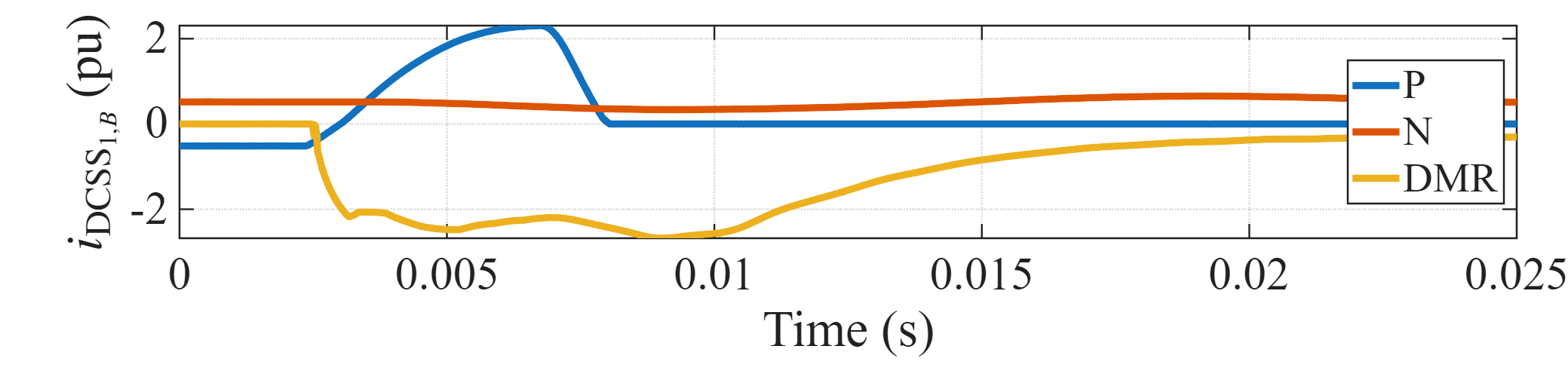}
    \caption{Time-domain responses of the positive-pole, negative-pole, and DMR DC voltages and currents, together with the protection signals, for a P-DMR fault applied at $80\%$ of subsea cable 1 measured from DC hub 1 at $t=1~\mathrm{ms}$ with a fault resistance of $10~\Omega$.}
    \label{fig:pdmr_cable_80}
\end{figure}

\begin{figure}[htbp]
    \centering
    \includegraphics[width=\linewidth]{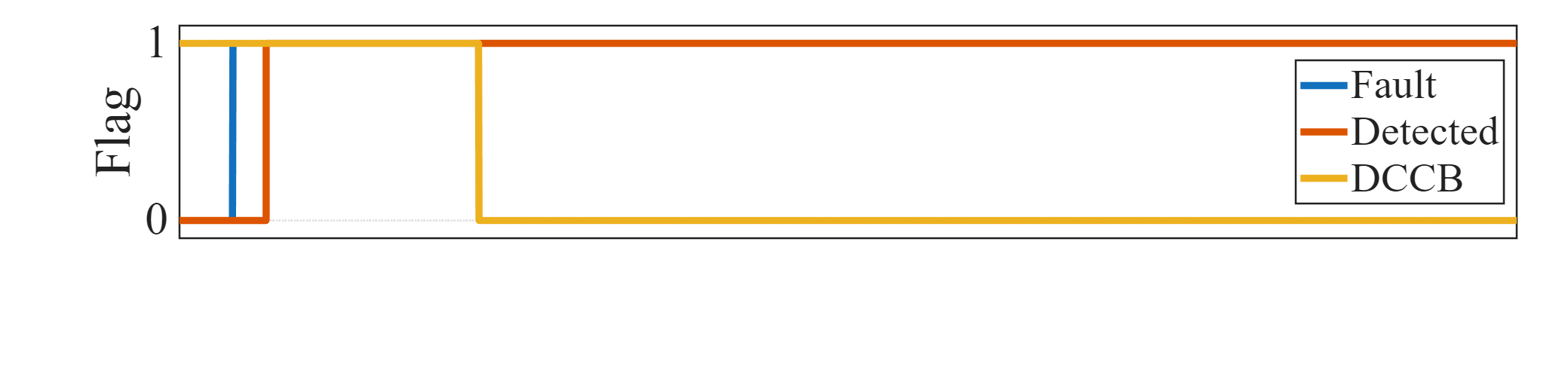} \\ [-0.5cm]
    \includegraphics[width=\linewidth]{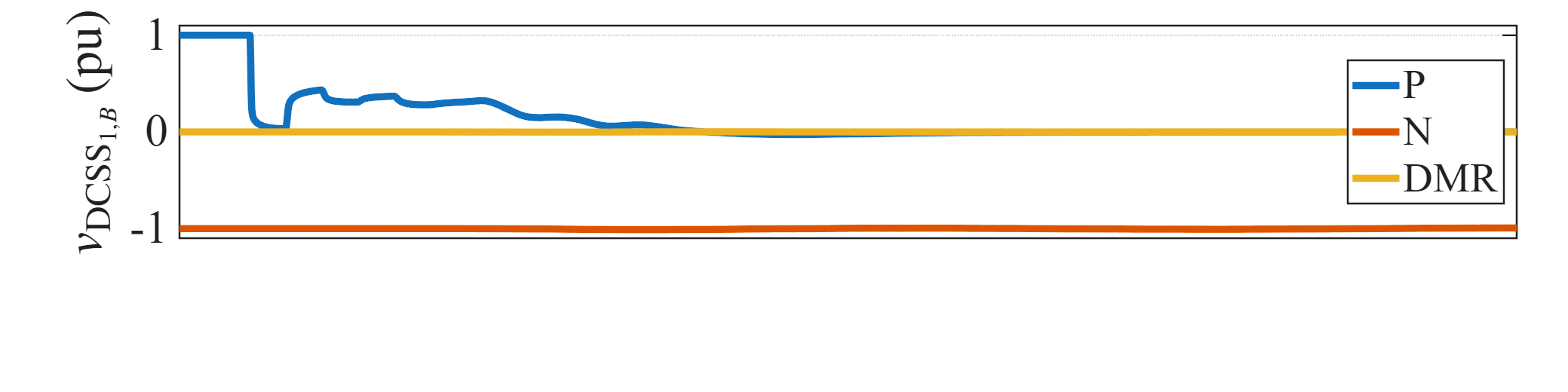} \\ [-0.5cm]
    \includegraphics[width=\linewidth]{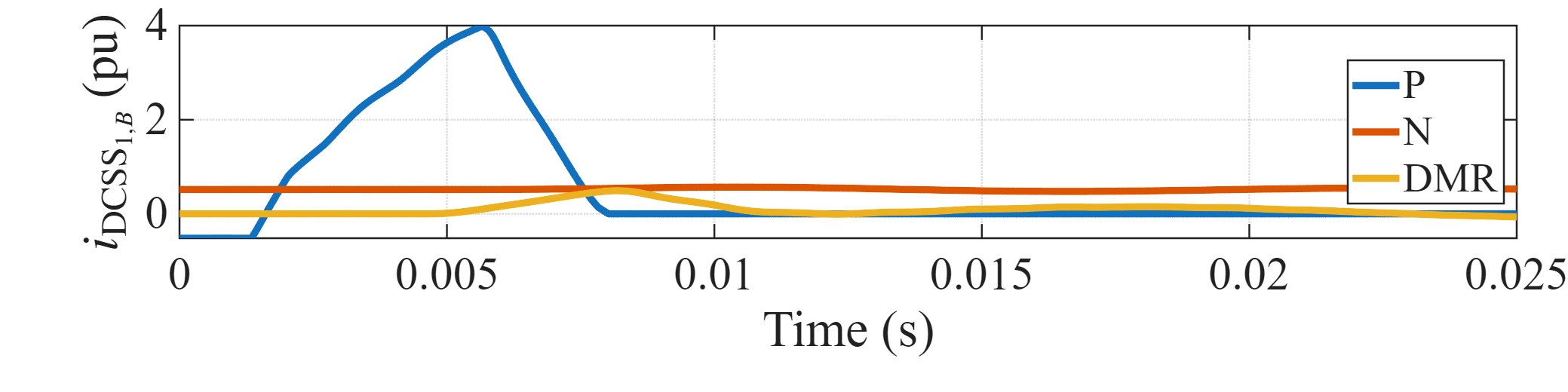}
    \caption{Time-domain responses of the positive-pole, negative-pole, and DMR DC voltages and currents, together with the protection signals, for a P-G fault applied at $20\%$ of subsea cable 1 measured from DC hub 1 at $t=1~\mathrm{ms}$ with a fault resistance of $10~\Omega$.}
    \label{fig:pg_cable_20}
\end{figure}

For the P-DMR fault, the responses in \cref{fig:pdmr_cable_80} show a relatively small change in DMR voltage, whereas the DMR current exhibits a steep transient following fault inception. This response is consistent with the DC breaker configuration: the positive and negative poles are equipped with DCCBs and CLRs, whereas the DMR has no corresponding CLR. Comparatively, the DMR current transient observed for the P-G fault in \cref{fig:pg_cable_20} is less pronounced, reflecting the differences between faults with and without DMR involvement. In both cases, the detection flag and trip command are asserted, indicating that an open command is issued to the DCCB.

\subsubsection{Influence of Fault Location} \label{subsubsec:cable_location}
\begin{figure}[htbp]
    \centering
    \includegraphics[width=\linewidth]{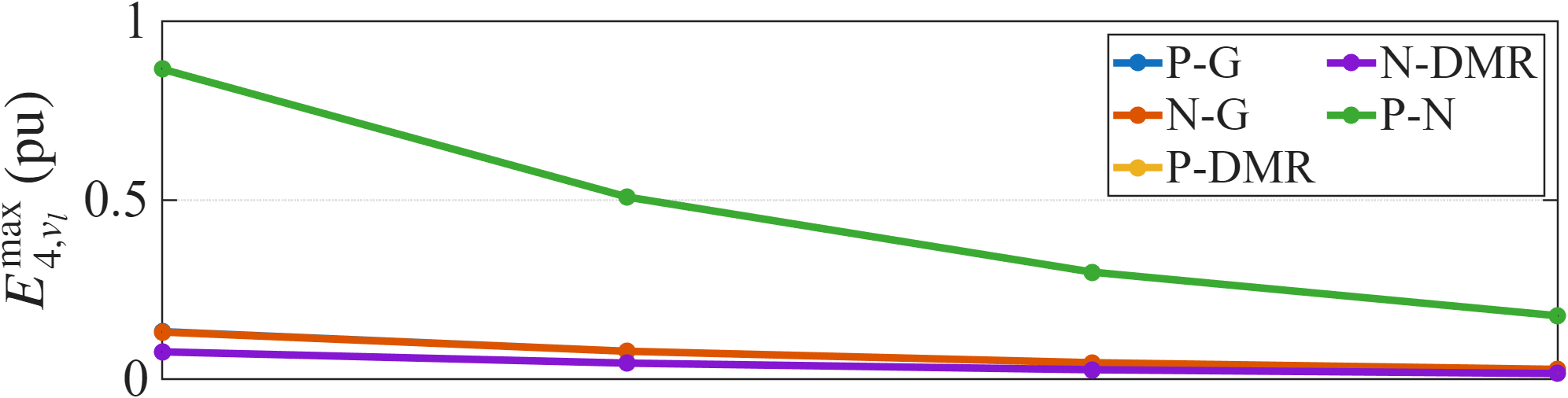} \\ [5pt]
    \includegraphics[width=\linewidth]{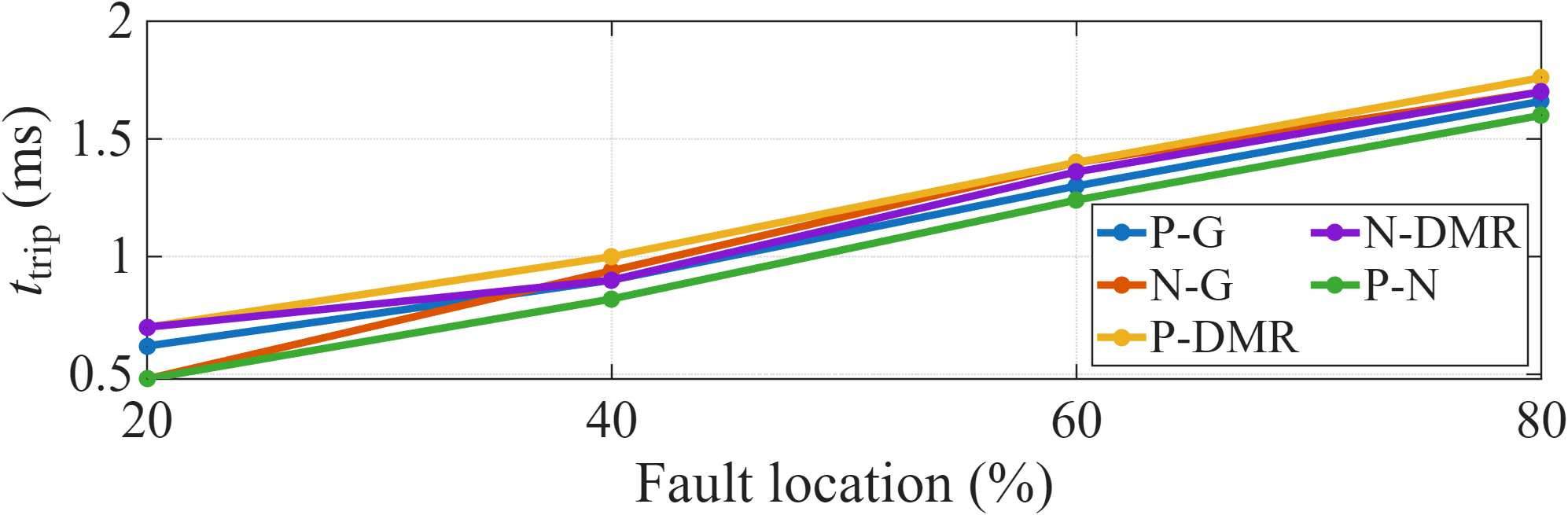}
    \caption{SWT-based protection performance for different cable fault types and locations at a fault resistance of $10~\Omega$, showing the maximum line-mode voltage detail energy and the detection time from fault inception to trip-signal assertion.}
    \label{fig:cable}
\end{figure}

The influence of fault location is assessed by comparing the maximum line-mode voltage detail energy and detection time across the cable-fault cases. The fault resistance is fixed at $10~\Omega$ for all cases, and the results for the tested fault types and locations are presented in \cref{fig:cable}. The maximum line-mode voltage detail energy generally decreases as the fault location moves farther from the measuring terminal, reflecting changes in the transient observed at the relay. In contrast, detection time generally increases with fault distance because the travelling wave requires more time to reach the measuring terminal. The measured detection time includes propagation, protection-processing, and communication delays. These results demonstrate the influence of fault location on the transient energy measured at the relay and the time required to assert the trip signal. The results across the tested locations and fault types, with resistance held constant, characterise the protection response under different cable-fault conditions.

\subsection{Busbar Fault Response} \label{subsec:busbar_fault_results}
Busbar-fault performance is assessed through time-domain responses and a comparative analysis of wavelet energy and detection time. The analysis considers the effects of fault type and resistance on the measured transients and the resulting protection response.

\subsubsection{Fault Transients and Protection Response} \label{subsubsec:busbar_waveforms}
\begin{figure}[htbp]
    \centering
    \includegraphics[width=\linewidth]{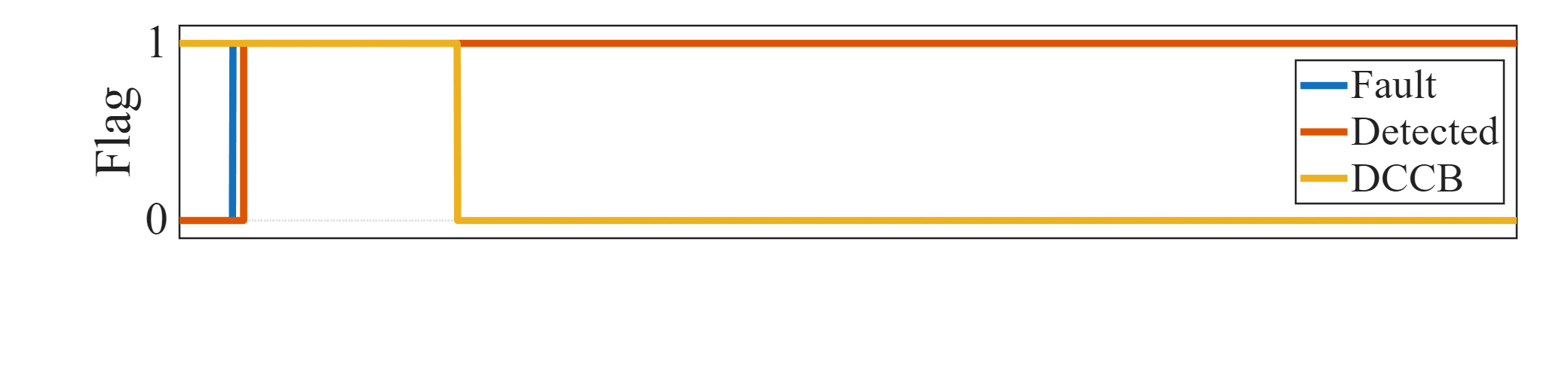} \\ [-0.5cm]
    \includegraphics[width=\linewidth]{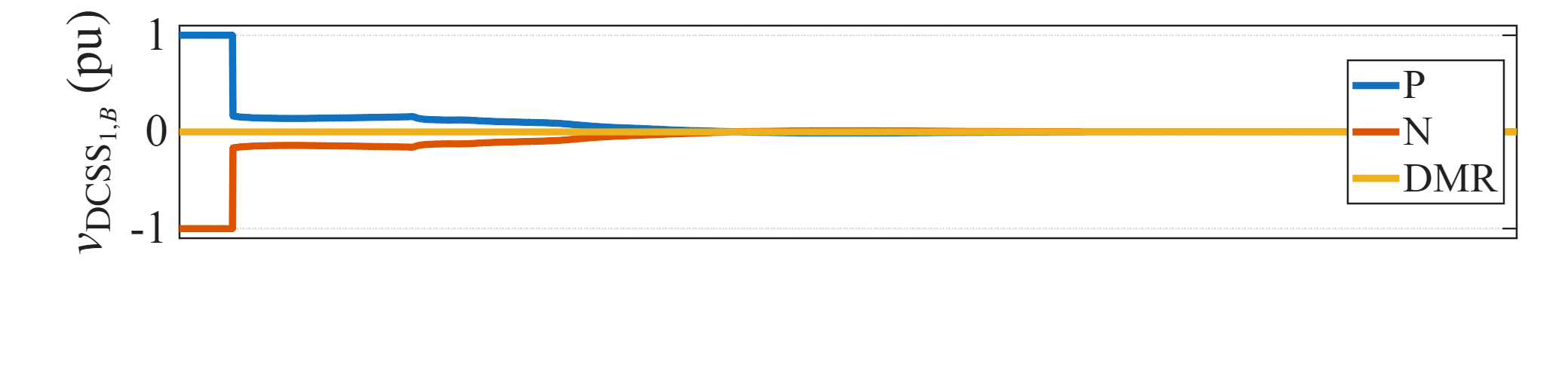} \\ [-0.5cm]
    \includegraphics[width=\linewidth]{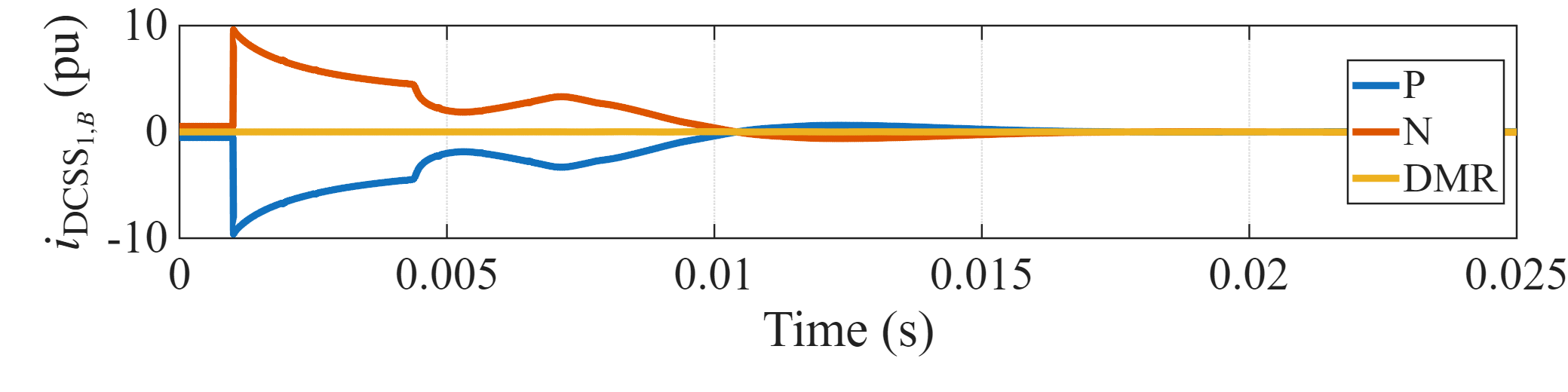}
    \caption{Time-domain responses of the positive-pole, negative-pole, and DMR DC voltages and currents, together with the protection signals, for a P-N fault applied at the busbar at the DC hub 1 end of subsea cable 1 at $t=1~\mathrm{ms}$ with a fault resistance of $10~\Omega$.}
    \label{fig:pn_busbar_fault}
\end{figure}

A P-N fault is applied at the busbar at the DC hub 1 end of subsea cable 1 at $t=1~\mathrm{ms}$ with a fault resistance of $10~\Omega$. The measured DC voltages and currents, fault-detection flag, breaker trip command, and breaker status are shown in \cref{fig:pn_busbar_fault}. The fault produces a sharp voltage transient and rapid changes in conductor currents. Following the disturbance, the detection flag is asserted and the trip command is issued. The subsequent change in breaker status demonstrates the isolation of the faulted section.

\subsubsection{Influence of Fault Type and Resistance}
\label{subsubsec:busbar_performance}

\begin{figure}[htbp]
    \centering
    \includegraphics[width=\linewidth]{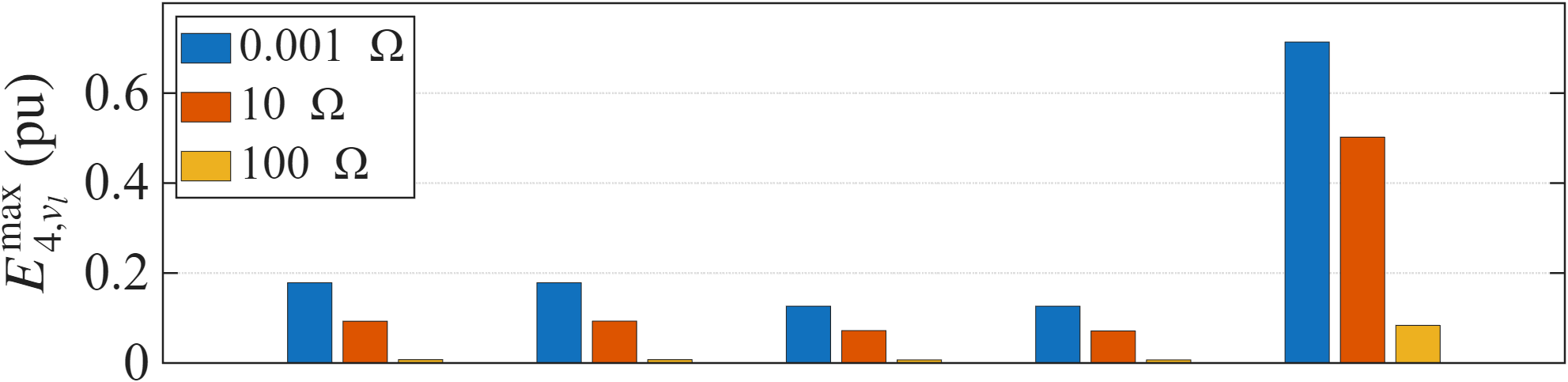} \\ [5pt]
    \includegraphics[width=\linewidth]{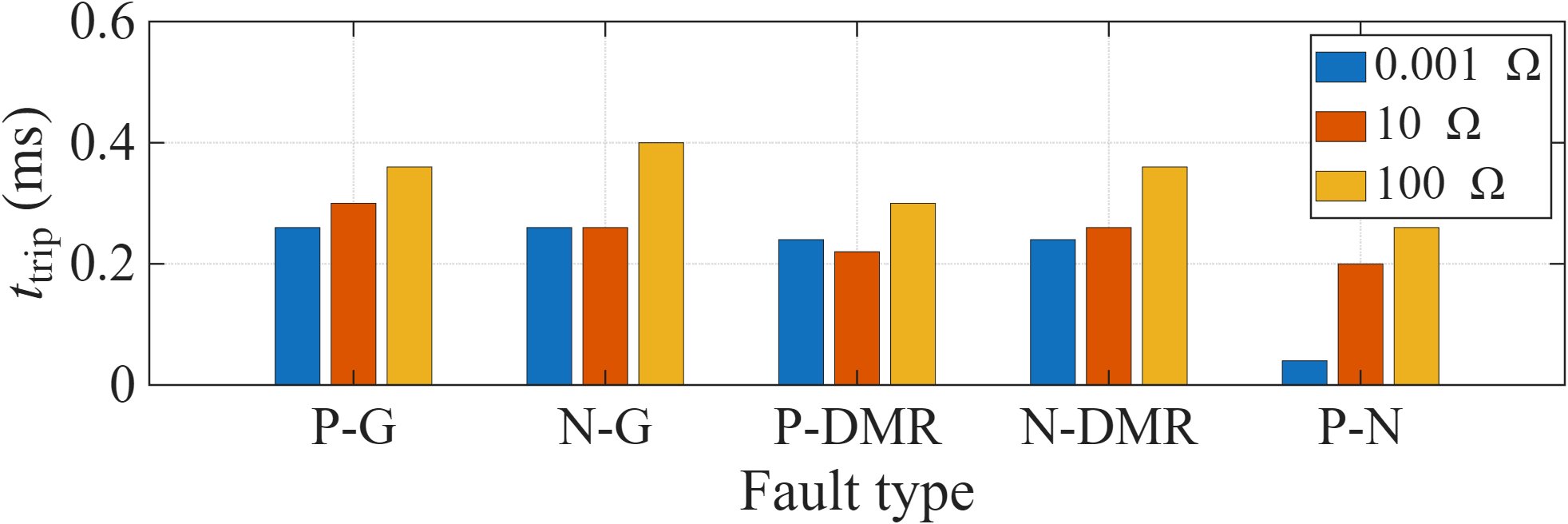}
    \caption{SWT-based protection performance for different busbar fault types and fault-resistance values, showing the maximum line-mode voltage detail energy and the detection time from fault inception to trip-signal assertion.}
    \label{fig:busbar}
\end{figure}

The maximum line-mode voltage, detail energy, and detection time are compared across busbar fault types and resistance values in \cref{fig:busbar}. The P-N fault produces the highest energy response. The P-G and N-G faults exhibit similar intermediate energy levels, whereas the P-DMR and N-DMR faults produce lower, comparable energy responses. The overlap between some fault types, particularly P-G and P-DMR, indicates that maximum line-mode voltage energy alone is insufficient for reliable classification. The classification method therefore combines conductor-specific detection signals derived from positive-pole voltage, negative-pole voltage, and DMR current, as described in \cref{subsec:fault_classification_results}.

The detection-time results in \cref{fig:busbar} quantify the interval from fault inception to the assertion of the trip signal. Differences across fault types and resistance values reflect variations in the measured transients and the time required for the detection criterion to be satisfied. 

\subsection{Fault Classification Performance} \label{subsec:fault_classification_results}
\begin{figure}[htbp]
    \centering
    \includegraphics[width=\linewidth]{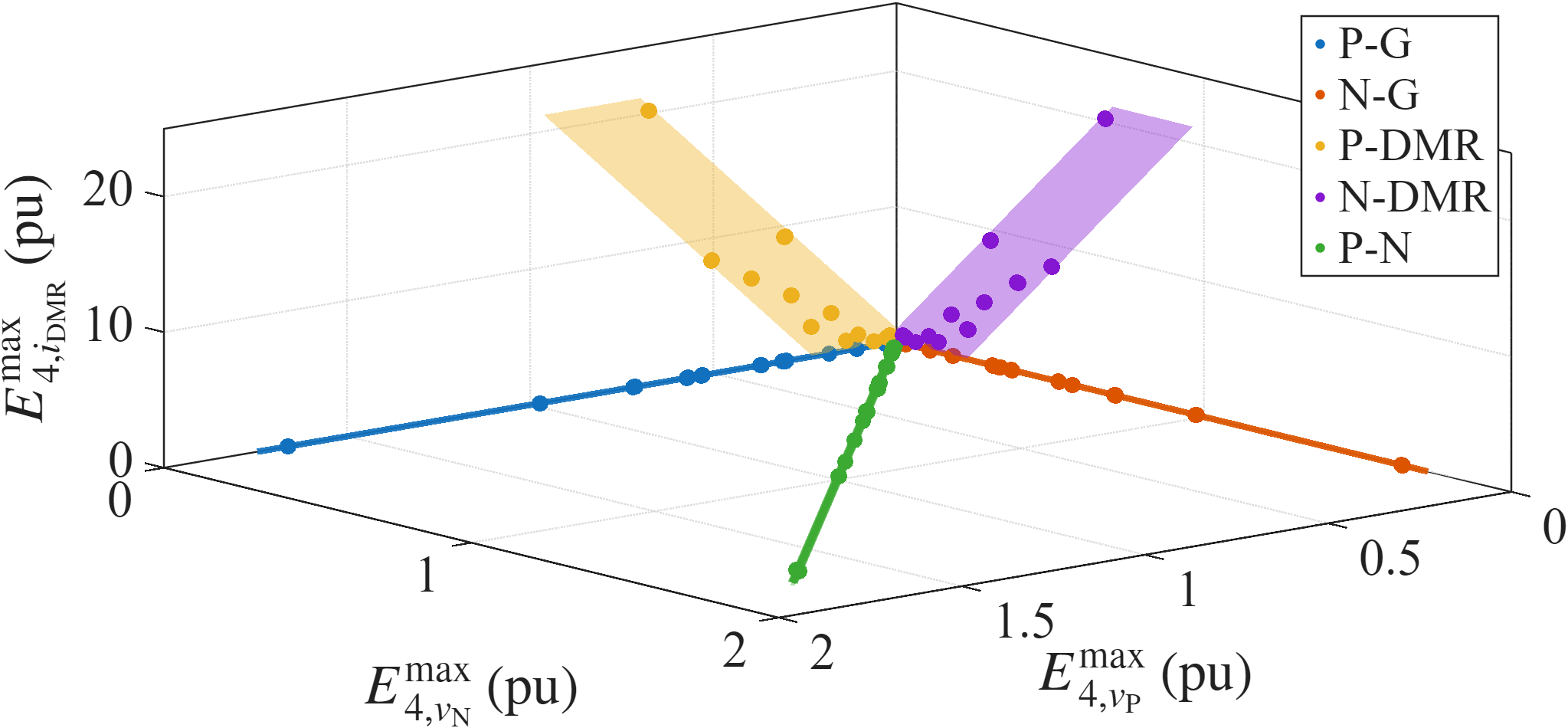}
    \caption{Three-dimensional distribution of the classification-feature energies for the tested fault types, locations, and resistances.}
    \label{fig:fault_classification_3d}
\end{figure}

The preceding results show that line-mode voltage energy varies with fault type, resistance, and location, but does not always uniquely distinguish faults with similar transient responses. Fault classification is therefore evaluated using three conductor-specific SWT energy features: positive-pole voltage energy, $E_{4,v_{\mathrm{P}}}$; negative-pole voltage energy, $E_{4,v_{\mathrm{N}}}$; and DMR current energy, $E_{4,i_{\mathrm{DMR}}}$. These features correspond to the signals processed by the three independent detection algorithms described in \cref{subsec:fault_classification}. Their joint distribution of their maximum values from each test case is shown in the three-dimensional scatter plot in \cref{fig:fault_classification_3d}.

The fault cases form distinct distributions in the three-dimensional feature space. P-G faults primarily extend along the positive-pole voltage-energy axis, whereas N-G faults primarily extend along the negative-pole voltage-energy axis. P-N faults exhibit elevated energy in both pole-voltage features, forming a distribution in the plane defined by $E_{4,v_{\mathrm{P}}}$ and $E_{4,v_{\mathrm{N}}}$. In contrast, P-DMR and N-DMR faults exhibit elevated DMR current energy together with the corresponding positive- or negative-pole voltage energy, respectively. These distributions incorporate the tested variations in fault location and resistance, allowing the separation between fault types to be examined under different fault conditions.

The distributions illustrate the distinct information provided by pole-voltage and DMR-current measurements. In particular, DMR current energy helps distinguish faults involving the DMR from pole-to-ground faults whose pole-voltage energy responses may be similar. In the implemented classifier, the moving-average SWT energy of each signal is compared with its predefined threshold to produce a binary detection output. The outputs are latched independently, and their combination determines the reported class according to \cref{eq:fault_classification}. When all three indicators are asserted, the classifier assigns the $\mathrm{Unknown}$ class because this combination does not uniquely identify one of the defined fault types.

\section{Conclusion} \label{sec:conclusion}
This paper presented an SWT-based fault-detection and classification method for a three-terminal bipolar MMC-HVDC system with a DMR and DCSSs incorporating interconnected DCCBs. The method uses line-mode voltage wavelet detail energy to detect fault-induced transients, while wavelet detail energy derived from positive-pole voltage, negative-pole voltage, and DMR current is used to classify fault types. Real-time HIL tests covering busbar and cable faults with varying fault types, resistances, and locations demonstrated the implementation of the method on embedded processing hardware and characterized the effects of fault location and travelling-wave propagation on transient energy and protection response time. The results show that pole-voltage and DMR-current detail energy features support the distinction between fault types that may exhibit similar voltage responses. Overall, the study demonstrates the feasibility of implementing and evaluating the proposed method in real time for the considered DC hub configuration. Future work will investigate its performance under a broader range of network operating conditions, measurement noise, and communication delays.

\section*{Acknowledgments}
The work is supported by Horizon Europe project PROSECCO, under grant agreement 101160687.

\section*{References}
\bibliographystyle{IEEEtran}
\bibliography{references}

\end{document}